\documentclass[sigplan,nonacm]{acmart}
\usepackage{listings}
\usepackage{algorithm}
\usepackage{algpseudocode}
\algrenewcommand\algorithmicindent{0.5em}
\usepackage{dashrule}
\usepackage{tabularx}  %
\usepackage{nicematrix}
\algtext*{EndFor}
\algtext*{EndIf}
\newcommand{\algsep}{\Statex\vspace{-0.55em}{\color{black!45}\hdashrule[0.35ex]{\linewidth}{0.4pt}{1.2pt}}\vspace{-0.35em}}
\usepackage[htt]{hyphenat} %
\newcommand{\claimRegsBeightLaunch}{32}                        %
\newcommand{\claimLowPlaneSixteen}{512}                        %
\newcommand{\claimPresetFlipWorst}{47\%}                       %

\newcommand{\claimBlockKiB}{31.25}                             %
\newcommand{\claimFourBlocksKiB}{125}                          %

\newcommand{\claimRegsBfour}{64}                               %
\newcommand{\claimBthreeRealMinTwelve}{952}                    %
\newcommand{\claimBthreeRealMaxTwelve}{1529}                   %
\newcommand{\claimBthreeRealMinSixteen}{814}                   %
\newcommand{\claimBthreeRealMaxSixteen}{1620}                  %
\newcommand{\claimFastestColumn}{Loghub \texttt{Windows}}      %
\newcommand{\claimFastestColumnPreset}{OnPair-16}              %
\newcommand{\claimFastestColumnRate}{1620}                     %
\newcommand{\claimFastestColumnDeRate}{650}                    %
\newcommand{\claimFastestColumnHopperRate}{1413}               %
\newcommand{\claimBeatsDeTrioPhrase}{all ten}                  %
\newcommand{\claimBeatsDeLfortySPhrase}{five}                  %
\newcommand{\claimShortLoghubLfortySPhrase}{all five}          %
\newcommand{\claimRealColumns}{10}                             %

\begin{document}
\title{FastPair: GPU-Optimized String Decoding}

\author{%
\href{mailto:joe@spiraldb.com}{Joseph Isaacs}\textsuperscript{1*},
\href{mailto:francesco.gargiulo@phd.unipi.it}{Francesco Gargiulo}\textsuperscript{1,2*},
\href{mailto:boncz@cwi.nl}{Peter Boncz}\textsuperscript{3},
\href{mailto:robert@spiraldb.com}{Robert Kruszewski}\textsuperscript{1},
\href{mailto:nick@spiraldb.com}{Nicholas Gates}\textsuperscript{1},\\
\href{mailto:rossano.venturini@unipi.it}{Rossano Venturini}\textsuperscript{2},
\href{mailto:will@spiraldb.com}{Will Manning}\textsuperscript{1},
\href{mailto:martin@spiraldb.com}{Martin Prammer}\textsuperscript{1\dag}%
}
\affiliation{%
  \institution{%
    \textsuperscript{1}Spiral, USA \quad
    \textsuperscript{2}University of Pisa, Italy \quad
    \textsuperscript{3}CWI, Netherlands
    }
  \country{}}
\thanks{\textsuperscript{*}Joseph Isaacs and Francesco Gargiulo contributed equally to this work.} 
\thanks{\dag Martin Prammer is the corresponding author (martin@spiraldb.com).}

\renewcommand{\shortauthors}{Isaacs, Gargiulo, et al.}

\begin{abstract}
Modern data systems compress data at rest and decompress it only when needed to preserve interconnect bandwidth. This design is often inefficient on GPU-based compute platforms because many conventional compression techniques exhibit serial data dependencies that limit GPU parallelism, leaving resources idle. Recent NVIDIA GPUs address this decoding deficiency through the Decompression Engine (DE), an on-die, fixed-function decompression accelerator for general-purpose compression formats such as Deflate, LZ4, and Snappy. Recent work has proposed string codecs that replace frequent substrings with fixed-width codes from a small, trained dictionary, making each code's lookup independent. While these lookups can run in parallel, the resulting scattered reads and short output writes still do not align well with GPU hardware, which handles contiguous memory accesses more efficiently. We present \textbf{FastPair}, a GPU decoder that optimizes the existing dictionary decoding process by reorganizing lookups and assembling decoded substrings for contiguous output writes. On a B300, FastPair decodes ten real-world columns 2.4 to 4.2$\times$ faster than the DE, reaching up to 1.6\,TB/s.

\end{abstract}

\maketitle
% overwrite author metadata
\hypersetup{
  pdfauthor={Joseph Isaacs, Francesco Gargiulo, Peter Boncz, Robert Kruszewski, Nicholas Gates, Rossano Venturini, Will Manning, Martin Prammer},
  pdfkeywords={GPU, string decoding, dictionary compression, FSST, OnPair, columnar storage},
  pdfpublisher={arXiv}}

\section{Introduction}
\label{sec:intro}

Data is typically stored in compressed form when not in use~\cite{snowflake,redshift,aurora}. When a data analytic query requires it, data must travel to the compute unit and be decompressed along the way. Decompressing at any point during the transfer sends the larger, decompressed form over one or more interconnects. Thus, decompressing on the device that performs the compute is often ideal.

However, decompressing on the computing device helps only if the decoder runs efficiently on that device. Widely used compression techniques were designed for CPUs with out-of-order, superscalar cores and often contain serial decoding dependencies. In contrast, GPUs rely on many independent threads to keep their execution units busy. Thus, serial decoding dependencies may leave a ported decoder with too little parallel work to use the device efficiently~\cite{sitaridi-gpu,codag,gsst}. The decoder's output rate then limits downstream operators that consume those bytes~\cite{cudf,sirius,crystal,heavydb}.

NVIDIA's Blackwell datacenter GPUs address this problem with the Decompression Engine (DE), a fixed-function hardware accelerator for Deflate, LZ4, and Snappy~\cite{nvcomp,nvcomp-de,blackwell-brief,blackwell-microbench}. The DE operates independently of the streaming multiprocessors (SMs). In this work, we investigate a software alternative that leverages a new class of compression techniques that expose enough independent work to use the SMs efficiently.

\begin{figure}[t]
\centering
\includegraphics[width=\columnwidth]{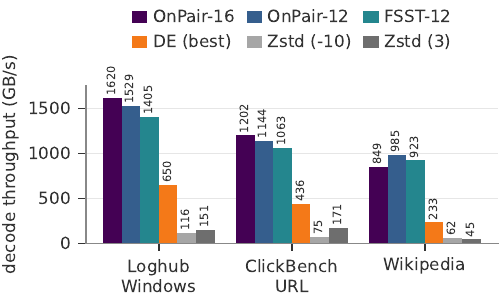}
\caption{Decode throughput on a B300 for the three codecs FastPair decodes, the Decompression Engine (DE) at its best codec and chunk size per column, and software nvCOMP-Zstd at compression levels $-10$ and $3$ (default frame size).}
\label{fig:teaser}
\end{figure}

\begin{table*}[t]
\centering
\caption{The five GPUs used throughout this work. All five devices carry a 65\,536-register file per SM.
}
\label{tab:arch}
\small
\begin{tabular}{@{}llrlrrrrr@{}}
\toprule
GPU & Architecture & CC & Mem. & Peak Mem. BW & SMs & \shortstack{Max Warps/SM} & \shortstack{SRAM/SM} & \shortstack{Max Shared Mem./SM} \\
\midrule
B300            & Blackwell Ultra    & 10.3 & HBM3e & 8.0\,TB/s  & 148 & 64 & 256\,KiB & 228\,KiB \\
H100-SXM 80\,GB & Hopper       & 9.0  & HBM3  & 3.35\,TB/s & 132 & 64 & 256\,KiB & 228\,KiB \\
A100-SXM 40\,GB & Ampere       & 8.0  & HBM2  & 1.56\,TB/s & 108 & 64 & 192\,KiB & 164\,KiB \\
RTX PRO 6000-SE & Blackwell    & 12.0 & GDDR7 & 1.60\,TB/s & 188 & 48 & 128\,KiB & 100\,KiB \\
L40S            & Ada Lovelace & 8.9  & GDDR6 & 0.86\,TB/s & 142 & 48 & 128\,KiB & 100\,KiB \\
\bottomrule
\end{tabular}
\end{table*}

Recent data management research has begun investigating alternatives to \emph{black-box} compression techniques, such as those the DE can decode. These works have proposed \emph{cascading} several data-specific \emph{lightweight} compression techniques~\cite{vortex,nvcomp-cascaded,f3,btrblocks,fastlanesff,anyblox}. These lightweight encodings preserve efficient access to individual values, allowing a query to retrieve a few values without decompressing an entire block~\cite{fastlanes,alp,fsst,onpair}. Together, these encodings can be applied iteratively, yielding compression ratios comparable to those of existing black-box techniques. Crucially, fine-grained accesses also enable parallel decoding.

We focus on FSST~\cite{fsst} and OnPair~\cite{onpair}, which compress strings by replacing frequent substrings with fixed-width \emph{codes}. Each code indexes a trained dictionary of short byte strings, called \emph{tokens}; decoding retrieves these tokens and concatenates them. These lookups can be performed in parallel because each lookup depends only on its code and the trained dictionary.

The main differences among these \emph{FSST-family} codecs stem from their dictionaries. For example, longer tokens can improve compression because a single code replaces more input bytes. However, storing longer tokens may require larger dictionaries. OnPair tokens can be up to 16 bytes, while FSST tokens can be up to 8 bytes. Although their encoders differ, these codecs share the same dictionary-lookup decoding process. Overall, dictionary lookups expose parallelism opportunities within and across strings, reducing serial dependencies that limit conventional GPU decoding. Thus, this family of compression codecs offers an opportunity to design an optimized GPU decoder.

However, this optimized decoder must still address several remaining challenges, such as organizing those lookups and their output for the GPU's memory system. Consecutive codes can refer to unrelated dictionary entries, while variable token lengths make output positions depend on all preceding token lengths. These scattered reads and short output writes fit poorly with hardware that most efficiently serves contiguous accesses from neighboring threads.

In this work, we present \textbf{FastPair}, a GPU decoder that optimizes dictionary decoding by reorganizing dictionary lookups and assembling decoded tokens for contiguous output writes. FastPair's design is based on the following three considerations:

First, OnPair's CPU decoder locates a token by reading its offset from a table and then fetching the token at that offset. Although lookups for different codes are independent, each lookup still requires two dependent reads. FastPair prepares fixed-size dictionary entries so that a thread can compute the token's address directly from the code, eliminating this dependency. Furthermore, the sixteen-byte entries needed for OnPair do not always need to be read in full, as many tokens contain eight bytes or fewer. FastPair splits the dictionary into low- and high-byte planes; this \emph{split read} optimization allows FastPair to read eight bytes per token, fetching the remaining bytes only for longer tokens (Section~\ref{sec:design}).

Next, writing these tokens in parallel requires knowing their output positions, which depend on the lengths of preceding tokens. FastPair precomputes output positions when the column is compressed, so that decoding does not have to recover them from the start of the code stream. It stores one position for each fixed-size group of codes, which we call a \emph{batch}. Threads assigned to a batch start at its stored position and sum token lengths in parallel to determine their individual destinations. This additional metadata, the \emph{sidecar}, adds about 1\% of data overhead to the stored column (Section~\ref{sec:design:placement}).

Finally, while knowing the output positions makes writes independent, it does not necessarily make them efficient, as each thread performs a small write. FastPair first assembles the tokens in a shared-memory buffer on the SM, then uses neighboring threads to copy contiguous parts of the completed batch to device memory. This optimization lets the hardware combine, or \emph{coalesce}, these contiguous writes into fewer memory accesses (Section~\ref{sec:design}).

\begin{table*}[t]
\centering
\caption{Dictionary statistics and compression ratios, per column and codec.
$\lvert Codes \rvert$ is the number of distinct codes that appear in the encoded column.
$\overline{Len}$ is the mean decoded bytes per code.
${\leq}8\mathrm{B}$ is the share of decoded codes whose token is eight bytes or fewer.
CR is the compression ratio; for baselines, $\mathrm{CR}_f$ is the CR of the fastest-decoding (B300) configuration, while $\mathrm{CR}\!\uparrow$ is the best CR found.
Real-world columns appear above the divider while synthetically generated columns are below.
}
\label{tab:datasets}
\small
\setlength{\tabcolsep}{1.8pt}
\setlength{\arrayrulewidth}{0.3pt}
\begin{NiceTabular}{@{}l<{\hspace{3.2pt}}I
>{\hspace{3.2pt}}rrrr<{\hspace{3.2pt}}I
>{\hspace{3.2pt}}rrrr<{\hspace{3.2pt}}I
>{\hspace{3.2pt}}rrr<{\hspace{3.2pt}}I
>{\hspace{3.2pt}}rr<{\hspace{3.2pt}}I
>{\hspace{3.2pt}}rr<{\hspace{3.2pt}}I
>{\hspace{3.2pt}}rr@{}}[custom-line={letter=I,dotted,color=black!60},xdots/shorten=0pt]
\toprule
 & \Block{1-4}{OnPair-16} &  &  &  & \Block{1-4}{OnPair-12} &  &  &  & \Block{1-3}{FSST-12} &  &  & \Block{1-2}{DE} &  & \Block{1-2}{Zstd} &  & \Block{1-2}{gANS} &  \\
\cmidrule(lr){2-5}\cmidrule(lr){6-9}\cmidrule(lr){10-12}\cmidrule(lr){13-14}\cmidrule(lr){15-16}\cmidrule(l){17-18}
(Dataset) Column & $\lvert Codes \rvert$ & $\overline{Len}$ & ${\leq}8\mathrm{B}$ & CR & $\lvert Codes \rvert$ & $\overline{Len}$ & ${\leq}8\mathrm{B}$ & CR & $\lvert Codes \rvert$ & $\overline{Len}$ & CR & $\mathrm{CR}_f$ & $\mathrm{CR}\!\uparrow$ & $\mathrm{CR}_f$ & $\mathrm{CR}\!\uparrow$ & $\mathrm{CR}_f$ & $\mathrm{CR}\!\uparrow$ \\
\midrule
FineWeb2 Mandarin & 65\,426 & 4.9 & 89\% & 2.5 & 4\,045 & 3.0 & 100\% & 2.0 & 4\,030 & 2.4 & 1.6 & 1.3 & 2.0 & 1.1 & 2.6 & 1.3 & 1.3 \\
Wikipedia & 65\,428 & 5.6 & 82\% & 2.8 & 4\,047 & 3.2 & 98\% & 2.1 & 3\,771 & 2.7 & 1.8 & 1.6 & 2.7 & 1.4 & 3.3 & 1.6 & 1.6 \\
CodeParrot & 65\,231 & 6.1 & 77\% & 3.0 & 4\,059 & 3.5 & 93\% & 2.3 & 3\,917 & 2.9 & 1.9 & 4.4 & 4.4 & 2.2 & 5.3 & 1.7 & 1.7 \\
Loghub \texttt{Android} & 61\,629 & 9.3 & 43\% & 4.6 & 3\,930 & 4.2 & 86\% & 2.8 & 3\,686 & 3.0 & 2.0 & 8.2 & 8.2 & 3.2 & 15.2 & 1.5 & 1.5 \\
ClickBench \texttt{URL} & 64\,428 & 8.6 & 51\% & 4.3 & 4\,012 & 4.7 & 81\% & 3.1 & 3\,678 & 3.2 & 2.1 & 6.5 & 6.5 & 6.0 & 9.1 & 1.4 & 1.4 \\
ClickBench \texttt{Title} & 64\,802 & 10.1 & 37\% & 5.0 & 4\,026 & 5.7 & 79\% & 3.8 & 2\,660 & 3.0 & 2.0 & 6.0 & 6.0 & 6.3 & 9.7 & 1.7 & 1.7 \\
Loghub \texttt{HDFS} & 63\,932 & 9.0 & 48\% & 4.5 & 3\,617 & 6.7 & 63\% & 4.4 & 1\,413 & 3.9 & 2.8 & 10.9 & 11.0 & 9.7 & 13.6 & 1.6 & 1.6 \\
Loghub \texttt{Thunderbird} & 59\,823 & 11.0 & 23\% & 5.5 & 3\,541 & 7.6 & 55\% & 5.0 & 2\,225 & 3.5 & 2.4 & 19.0 & 19.0 & 5.1 & 26.5 & 1.5 & 1.5 \\
Loghub \texttt{Spark} & 62\,485 & 11.2 & 21\% & 5.5 & 3\,640 & 8.3 & 42\% & 5.5 & 2\,037 & 4.0 & 2.9 & 15.4 & 15.5 & 15.4 & 19.8 & 1.6 & 1.6 \\
Loghub \texttt{Windows} & 37\,853 & 12.1 & 18\% & 6.0 & 3\,551 & 10.1 & 37\% & 6.7 & 1\,793 & 4.5 & 3.3 & 9.9 & 18.3 & 18.7 & 137.7 & 1.6 & 1.6 \\
\midrule
TPC-H \texttt{c\_address} & 65\,180 & 2.2 & 100\% & 1.1 & 3\,904 & 1.9 & 100\% & 1.3 & 3\,904 & 1.9 & 1.3 & 1.0 & 1.3 & 0.9 & 1.2 & 1.3 & 1.3 \\
TPC-H \texttt{l\_comment} & 64\,971 & 10.4 & 33\% & 5.2 & 3\,862 & 7.7 & 58\% & 5.1 & 1\,044 & 4.0 & 3.1 & 4.6 & 4.6 & 1.9 & 4.2 & 1.9 & 1.9 \\
TPC-H \texttt{o\_clerk} & 45\,000 & 15.0 & 0\% & 7.5 & 3\,837 & 7.7 & 48\% & 5.1 & 783 & 3.8 & 3.0 & 2.8 & 5.3 & 6.7 & 6.8 & 2.2 & 2.3 \\
TPC-H \texttt{l\_shipinstruct} & 5 & 9.6 & 40\% & 25.2 & 5 & 9.6 & 40\% & 25.2 & 8 & 6.0 & 15.8 & 25.0 & 25.2 & 24.3 & 28.1 & 2.1 & 2.1 \\
TPC-H \texttt{ps\_comment} & 65\,049 & 12.7 & 16\% & 6.3 & 3\,865 & 10.3 & 32\% & 6.8 & 936 & 4.9 & 3.9 & 5.8 & 5.8 & 4.7 & 6.4 & 1.9 & 1.9 \\
\bottomrule
\end{NiceTabular}
\end{table*}

Across the ten real-world data columns we explore, FastPair decodes at 2.4 to 4.2$\times$ the DE's rate on a B300. It reaches 1.6\,TB/s when decoding Loghub \texttt{Windows} with OnPair-16, 2.5$\times$ the DE's rate (Figure~\ref{fig:teaser}). We perform two studies to better understand FastPair's decoding performance:

First, we conduct a sensitivity study to understand how FastPair's design choices interact with the input data and the GPU's on-chip resources (Section~\ref{sec:mb}). Per-thread overheads can be better amortized by assigning each thread more codes to process; however, this optimization requires more registers and shared memory, which can limit parallelism. The token-length distribution also matters: split reads help when short tokens are common, whereas reading the full entry at once avoids a second read for long tokens.

Then, we use hardware counters to examine how dictionary reads and decoded output assembly use the GPU's memory system. While FastPair coalesces writes to device memory, assembling the output still requires many short shared-memory writes. Together with the scattered dictionary reads, these writes can saturate the L1 cache's access pipeline while device-memory bandwidth remains available (Section~\ref{sec:eval:bound}). Consistent with this on-chip limit, decoding on the three GPUs with high-bandwidth memory (HBM) scales nearly linearly with SM clock speed across the evaluated columns while memory clocks remain fixed (Section~\ref{sec:eval:arch}).

The work is structured as follows: Section~\ref{sec:background} describes the codecs and GPU mechanisms FastPair uses. Then, Section~\ref{sec:design} presents the FastPair decoder. We explore and measure FastPair's design choices in Section~\ref{sec:mb}. Section~\ref{sec:evaluation} evaluates FastPair and examines how FastPair uses GPU resources. Section~\ref{sec:related} discusses prior work. Finally, we conclude and discuss future work in Section~\ref{sec:conclusion}.

\section{Background}
\label{sec:background}

In this section, we discuss the string codecs FastPair decodes, including how their dictionaries are built (Section~\ref{sec:bg:codecs}) and OnPair's original CPU-optimized decoder (Section~\ref{sec:bg:cpu}). Then, we discuss the architectural components that enable high-performance parallel decoders on GPUs (Section~\ref{sec:bg:gpu}).

\subsection{Dictionary-Compressed Strings}
\label{sec:bg:codecs}

Dictionary encodings and their use for compression are long established~\cite{storer-szymanski}. Their re-emergence is part of a broader trend in data management research exploring lightweight compression techniques and their interactions with data analysis tasks~\cite{alp,galp,anema-gpu-fsst,gsst,btrblocks,vortex,f3,optfsst,gpucompressedsql}. These techniques enable random access to compressed data, eliminating the need to decompress large blocks to extract scattered values. Random access to values also enables highly parallel decoding, which generally aligns with heterogeneous data management platforms such as GPU-based data analytic engines~\cite{theseus,sirius,heavydb}.

In this work, we focus on lightweight string codecs. Older formats derive the dictionary from the text itself: LZW extends it while scanning~\cite{welch}, and LZ77 replaces repeated segments with back-references to earlier positions~\cite{ziv-lempel}. Thus, both techniques require decoding a block from its start to recover any value. In contrast, the two lightweight string codecs we explore in this work, FSST~\cite{fsst} and OnPair~\cite{onpair}, train a dictionary whose entries remain fixed during decoding. These \emph{FSST-family} techniques rewrite each row as dictionary \emph{codes}, where each code indexes a stored byte-string \emph{token}; given this dictionary, each dictionary code can be looked up without decoding the preceding text.

\emph{Cascaded} encoding techniques apply a sequence of lightweight encodings~\cite{vortex,nvcomp-cascaded,f3,btrblocks,fastlanesff,anyblox}. FSST-family encodings particularly benefit from these optimizations, as they encode string data into an integer form that can then benefit from a variety of numeric-specific lightweight encodings~\cite{fastlanes,fastlanesgpu,galp,shanbhag2022tile,l3,alp}.

Table~\ref{tab:datasets} lists the columns from each dataset we explore in this work. Their compression-related statistics follow from three differences in dictionary construction: a codec's \emph{training procedure}, \emph{code width}, and \emph{maximum token length}.

FSST and OnPair use different dictionary training methods. FSST trains over several passes on a sample, concatenating frequent adjacent code pairs into longer tokens and keeping those that cover the most sample bytes, a process that fills small dictionaries well. OnPair instead merges frequent adjacent substrings in a single sequential pass, following byte-pair encoding~\cite{gage} and its relative Re-Pair~\cite{larsson-moffat}, but merging as it scans rather than tracking where each pair occurs. Because a merge joins two entries already in the dictionary, tokens lengthen as training proceeds.

FSST-family encoding techniques use fixed-width codes, so the code width (or maximum dictionary cardinality) determines how many entries a dictionary can address. The maximum token length sets an upper bound on how many bytes a single code may expand to. In this work, we evaluate OnPair-16, OnPair-12, and FSST-12, where the suffix indicates the code width: -16 and -12 address up to $65\,536$ and $4\,096$ entries, respectively. FSST-12 caps the maximum token length at 8 bytes, while OnPair-16 and OnPair-12 cap it at 16 bytes. These differences are visible in Table~\ref{tab:datasets}: At the same code width, OnPair-12's mean token length is longer than FSST-12's in every column except \texttt{c\_address}, whose values are random characters.

Notably, none of these three techniques requires \emph{escape codes}: markers for literals the dictionary cannot represent. Both FSST-12 and OnPair-16/-12 place all one-byte literals in their dictionaries, eliminating the need for such codes. FSST-8 and its 255-entry dictionary cannot include all one-byte literals; instead, any missing byte is stored immediately after the reserved escape code. A decoder starting at an arbitrary input position must therefore determine whether the starting byte is a dictionary code or a literal following an escape. Without escapes, threads can begin dictionary lookups at any fixed-width code boundary without inspecting preceding input. This allows the input to be divided into fixed-size groups of codes for parallel processing. However, fixed-size groups of codes can expand to different numbers of bytes. Thus, writing their output into a single packed buffer requires knowing the total decoded length of the preceding groups.

\begin{lstlisting}[language={},float=tbp,captionpos=t,caption={OnPair's CPU decoding loop.},label={lst:cpu-decoder}]
let c     = *codes_ptr.add(i + k) as usize;
let entry = *table_ptr.add(c);  // u64 = (off<<16)|len
let off   = (entry >> 16) as usize;
let len   = (entry & 0xffff) as usize;
std::ptr::copy_nonoverlapping(
    dict_ptr.add(off),
    cursor,
    crate::MAX_TOKEN_SIZE,  // always 16
);
cursor = cursor.add(len);  // advance by true len
\end{lstlisting}

\begin{table}[t]
\centering
\caption{Terms used throughout this paper. Some definitions conflict across vendors; we use NVIDIA terminology.}
\label{tab:terms}
\small
\begin{tabularx}{\columnwidth}{@{}l X@{}}
\toprule
Term & Definition \\
\midrule
SM & Streaming multiprocessor: an execution unit with its own registers, scratchpad, and L1.  \\
register & Per-thread storage allocated from each SM. \\
thread & The unit that executes the kernel body. \\
thread block & A group of threads assigned to the same SM. \\
shared memory & Scratchpad on each SM, shared within a block. \\
carveout & The split of an SM's single SRAM array between shared memory and L1, requested by the kernel. \\
\midrule
warp & A group of 32 threads executing common instructions (i.e., single-instruction, multiple-thread). \\
lane & A thread's position within its warp. \\
resident blocks & Blocks an SM holds at once. \\
launch bounds & Compile-time parameters: threads per block ($T$) and requested blocks per SM ($B$). Together, they bound a kernel's register allocation. \\
\midrule
device memory & GPU DRAM. \\ %
global memory & The address space every thread can reach. \\
sector & An aligned 32-byte portion of a cache line. A 128-byte L1 cache line contains four sectors. \\
wavefront & The unit of work the L1 pipe processes in one cycle. One warp's memory instruction becomes one or more wavefronts~\cite{ncu}. \\
\bottomrule
\end{tabularx}
\end{table}

\subsection{CPU-based String Decoding}
\label{sec:bg:cpu}

In this section, we examine OnPair's CPU decoder (Listing~\ref{lst:cpu-decoder}).

OnPair's decoder uses a four-way-unrolled loop. It reads a code and its packed offset and length, copies sixteen bytes, and advances the output cursor by the true token length. Fetching the token requires two dependent reads: the table entry supplies the address for the token load. Note that this loop is already optimized for CPUs: copying a constant sixteen bytes and advancing by the true length is branch-free. Further, this sequence lowers to one unaligned 128-bit store on x86-64 and AArch64. Finally, because tokens are written in stream order, the next write overwrites the excess bytes from the previous copy.

\subsection{GPU Architecture}
\label{sec:bg:gpu}

\begin{algorithm}[t]
\caption{The FastPair Decoding Algorithm}
\label{alg:stageddrain}
\begin{algorithmic}[1]
\Require lane $\ell \in [0,32)$, $32 \cdot K$ codes, output $\mathit{out}$ at \texttt{cursor}
\Require dictionary planes $\mathit{dict}_{\mathrm{lo}}$, $\mathit{dict}_{\mathrm{hi}}$; token lengths $\mathit{lens}$
\Require $K$, tokens per thread
\Require $W$, token bytes in $\mathit{dict}_{\mathrm{lo}}$
\Require $H$, rounds of queued reads held across the emit
\Require $|c|,|\mathit{len}|,|\mathit{off}|,|\mathit{lo}|=K$; $|\mathit{hi}|=H$ per lane
\algsep
\For{$k \gets 0 \ldots K-1$}\Comment{\textbf{input}, \textbf{gather}}
  \State $c_k \gets \mathit{codes}[\ell + 32k]$
  \State $\mathit{len}_k \gets \mathit{lens}[c_k]$
  \State $\mathit{lo}_k \gets \mathit{dict}_{\mathrm{lo}}[c_k]$\Comment{$W$ bytes}
\EndFor
\algsep
\State $\mathit{off}, \mathit{total} \gets \Call{WarpPrefixSum}{\mathit{len}}$\Comment{\textbf{scan}}
\State $\mathit{shift} \gets \texttt{cursor} \bmod 16$
\algsep
\State $n \gets 0$\Comment{\textbf{queue}}
\For{$k \gets 0 \ldots K-1$}\label{alg:line:compact}
  \State $m \gets \Call{Ballot}{\mathit{len}_k > W}$
  \If{$\mathit{len}_k > W$}
    \State $\mathit{queue}[n + \Call{Rank}{m, \ell}] \gets (c_k,\ \mathit{off}_k{+}W,\ \mathit{len}_k{-}W)$
  \EndIf
  \State $n \gets n + \Call{PopCount}{m}$
\EndFor
\State \Call{SyncWarp}{}
\algsep
\For{$r \gets 0 \ldots H-1$}\label{alg:line:hoist}\Comment{\textbf{hoist}}
  \If{$\ell + 32r < n$}
    \State $\mathit{hi}_r \gets \mathit{dict}_{\mathrm{hi}}[\mathit{queue}[\ell + 32r].c]$ \Comment{$\mathit{dict}_{\mathrm{hi}}$ reads}
  \EndIf
\EndFor
\algsep
\For{$k \gets 0 \ldots K-1$, $j \gets 0 \ldots W-1$}\Comment{\textbf{emit}}
  \If{$j < \min(\mathit{len}_k, W)$}
    \State $\mathit{scratch}[\mathit{shift} + \mathit{off}_k + j] \gets \mathit{lo}_k[j]$
  \EndIf
\EndFor
\For{$r \gets 0 \ldots K-1$}
  \If{$\ell + 32r < n$}
    \State $q \gets \mathit{queue}[\ell + 32r]$
    \State $v \gets \mathit{hi}_r$ \textbf{if} $r < H$ \textbf{else} $\mathit{dict}_{\mathrm{hi}}[q.c]$
    \For{$j \gets 0 \ldots q.\mathit{len}-1$}
      \State $\mathit{scratch}[\mathit{shift} + q.\mathit{off} + j] \gets v[j]$
    \EndFor
  \EndIf
\EndFor
\algsep
\State \Call{SyncWarp}{}\Comment{\textbf{drain}}
\State $\mathit{head} \gets \min((16 - \mathit{shift}) \bmod 16,\ \mathit{total})$
\State $\mathit{body} \gets \lfloor (\mathit{total} - \mathit{head})/16 \rfloor$
\State $t \gets \mathit{head} + 16\,\mathit{body}$
\If{$\ell < \mathit{head}$} \Comment{head}
\State $\mathit{out}[\texttt{cursor} + \ell] \gets \mathit{scratch}[\mathit{shift} + \ell]$
\EndIf
\For{$k \gets \ell$ \textbf{to} $\mathit{body} - 1$ \textbf{step} $32$}\Comment{aligned body}
  \State $\mathit{out}[\texttt{cursor} + \mathit{head} + 16k] \gets \mathit{scratch}[\mathit{shift} + \mathit{head} + 16k]$ %
\EndFor
\If{$\ell < \mathit{total} - t$}\Comment{tail}
\State $\mathit{out}[\texttt{cursor} + t + \ell] \gets \mathit{scratch}[\mathit{shift} + t + \ell]$ 
\EndIf
\State $\texttt{cursor} \gets \texttt{cursor} + \mathit{total}$\Comment{next batch}
\end{algorithmic}
\end{algorithm}

Because GPU terminology is often vendor-specific, we define the terms used throughout this work in Table~\ref{tab:terms}. Table~\ref{tab:arch} identifies the capabilities of the devices used in this work. 

A GPU delivers data to its streaming multiprocessors (SMs) via device memory, a GPU-wide L2 cache, and per-SM SRAM, which serves as the L1 cache and \emph{shared memory} (a scratchpad). A kernel can request that SRAM be allocated as shared memory rather than L1, known as the \emph{carveout}. Unallocated SRAM is used for the L1 cache~\cite{cudaguide}. Larger shared-memory allocations reduce the L1 space available for dictionary reads.

An SM hides memory latency by keeping several warps resident and switching between them. The resident warp count relative to the hardware maximum is known as its \emph{occupancy}. Warps are assigned to an SM in groups called \emph{thread blocks} (blocks). Each block reserves registers and shared memory; thus, larger allocations can leave room for fewer blocks and, consequently, fewer resident warps.

However, raising occupancy is not always beneficial. For example, \emph{coarsening}, the process of assigning additional work to each thread, can improve performance~\cite{volkov}. While coarsening gives each thread more work and uses more per-thread resources (such as registers), lowering occupancy, it better amortizes per-thread overheads.

While keeping more work in flight helps hide memory latency, memory-access throughput is limited by how quickly the L1 pipe can process requests. A warp's global-memory instruction produces a single request containing the addresses requested by all participating lanes. The hardware divides this request into \emph{wavefronts}, where a pipeline stage processes one wavefront per cycle. Requests requiring several wavefronts consume more of the L1 pipe's processing capacity, even if all their data is cached. Shared-memory reads and writes also consume the L1 pipe's processing capacity~\cite{ncu}.

We distinguish the L1 pipe's \emph{access rate}, measured in wavefronts per cycle, from the cache's \emph{byte bandwidth}, measured in bytes per cycle. When the 32 lanes of a warp read contiguous, aligned addresses, GPU hardware can merge the lanes' reads into a few wide accesses. When those same lanes read 32 unrelated addresses, their reads may span 32 different sectors, even though each lane needs only a few bytes. Depending on the addresses and access width, serving the same number of sectors may require one wavefront or several~\cite{ncu}. These scattered accesses can exhaust the cache's access rate while its byte bandwidth remains available.

To output decoded values, the threads must also recover the positions the CPU decoder obtains by advancing its cursor. A warp's \emph{parallel prefix sum} computes these positions by summing token lengths across lanes, using a handful of warp shuffles rather than a serial pass over the tokens (Section~\ref{sec:design:scan}).

\begin{table}[t]
\centering
\caption{Definitions of warp primitives used in Algorithm~\ref{alg:stageddrain}.}
\label{tab:warp-primitives}
\small
\begin{tabularx}{\columnwidth}{@{}lX@{}}
\toprule
Primitive & Definition \\
\midrule
\textsc{Ballot}$(p)$ & Returns a 32-bit mask whose bit $\ell$ is set when lane $\ell$'s predicate $p$ is true. \\
\textsc{PopCount}$(m)$ & Counts the set bits in mask $m$. \\
\textsc{Rank}$(m,\ell)$ & Counts the set bits in $m$ below bit $\ell$, giving that lane's position in the queue. \\
\textsc{WarpPrefixSum}$(\mathit{len})$ & Returns exclusive byte offsets for all $32K$ tokens in stream order, $\ell+32k$, and their total decoded length. \\
\textsc{SyncWarp} & Waits for all 32 lanes and orders their memory accesses so prior shared-memory writes are visible to subsequent reads. \\
\bottomrule
\end{tabularx}
\end{table}

\section{Design}
\label{sec:design}

Algorithm~\ref{alg:stageddrain} presents the FastPair decoder and its stages, which use warp primitives from Table~\ref{tab:warp-primitives}.
Decoding a dictionary-compressed string requires reading a code, looking up its token, and appending the token's bytes to the output. This process has multiple properties that make parallel decoding difficult. First, output placement depends on all previous tokens because tokens are concatenated without padding. Second, locating a token requires two dependent reads: the offset table must be read before the token's bytes. Third, each token is written to the output in a short, unaligned copy.

To address these challenging properties, FastPair uses precomputed offsets and a parallel prefix sum to resolve output placement. Further, its fixed-stride dictionary layout makes each token directly addressable from its code, while staging coalesces writes to device memory. However, these optimizations consume other contested resources, creating trade-offs that we explore in Section~\ref{sec:design:model}.

\subsection{Design Overview}
\label{sec:design:placement}

FastPair distributes decoding across many GPU threads, each handling a small portion of the code stream. Within a warp, threads cooperate on a batch of consecutive codes. Each thread processes $K$ codes, resulting in each warp processing $32 \cdot K$ codes per batch.

Each warp reads its batch's output position (\texttt{cursor}) from the sidecar before decoding. The sidecar stores each batch's starting location. By recording this information at compression time, each warp can start without waiting for earlier batches. Including the sidecar in the compression output adds about 1\% of storage overhead. We observe that, for the columns explored in this work, regenerating the sidecar at decode time instead results in about 20\% slower decoding, and up to 35\% on short-token columns. The decoding penalty tends to be larger for columns with shorter tokens because regeneration processes more codes per decoded byte.

Within a batch, threads fetch and decode their tokens independently, using a parallel prefix sum of token lengths to determine each token's local output position. Each thread writes its decoded output to a shared-memory buffer, enabling neighboring threads to copy contiguous segments to device memory via coalesced writes.

\subsection{Decode Stages}

In this section, we walk through the FastPair decoder stages step by step. These stages are labeled in Algorithm~\ref{alg:stageddrain}.

The first stage is the \textbf{input} stage, where lanes read consecutive codes in $K$ rounds, keeping input accesses coalesced. This input feeds the \textbf{gather} stage, where each thread reads the length and the first $W$ bytes of each of its $K$ tokens. Because each token is directly addressable from its code (an index into a fixed-stride dictionary), FastPair does not require a dictionary offset table. The dictionary is stored as two planes: a \emph{low plane} holding $W$ bytes per entry, read for every token, and a \emph{high plane} holding the remaining bytes (also at a fixed stride), read only for tokens longer than $W$. %
For OnPair, $W=8$ induces split reads, while $W=16$ fetches the full dictionary entry at once. Token lengths are packed into nibbles in a separate table, storing each token's length minus one. As the length read and the token read depend only on the original code, they can proceed independently.

After the gather-stage reads finish, a warp \textbf{scan}\label{sec:design:scan} resolves the batch-local offsets for each token by summing the lengths of preceding codes. Adding these offsets to \texttt{cursor} gives the final output positions.

If any tokens longer than $W$ bytes are present, an additional lookup is \textbf{queued} for the remaining bytes, allowing the warp to redistribute this extra work across its lanes. Within each input round, a warp \textsc{Ballot} identifies the lanes holding long tokens; a prefix count assigns their requests to consecutive queue slots. The warp processes the completed queue in rounds of up to 32 requests, one per lane. Before outputting the batch, the warp \textbf{hoists} the first $H$ rounds of queued reads, so those loads are in flight before the emit begins. 

Then, each thread \textbf{emits} the first $W$ bytes of each token to a shared-memory staging buffer ($\mathit{scratch}$), or the entire token if it is shorter. The scanned offsets assign each token a disjoint range, so these writes can proceed in parallel without the overlapping copies used by the CPU decoder. Threads handling queued reads fill the remaining bytes of long tokens.

Once the batch is fully staged, the warp \textbf{drains} the buffer to global memory. Neighboring lanes copy consecutive, aligned 16-byte chunks, so most of the output uses coalesced stores. To align these copies, the first token is placed at a scratch-buffer offset of \texttt{cursor} modulo 16, aligning the buffer with the destination. Any unaligned head and trailing bytes are copied individually, so no write extends into another batch.

\subsection{Performance Modeling}
\label{sec:design:model}

FastPair's batch size determines how much work each warp performs before draining its output. Larger batches amortize the scan and drain overheads, but also require more registers and shared memory. Hoisting introduces a similar trade-off: keeping more reads in flight requires registers to hold their results. In this section, we model these resource requirements to understand how $K$ and $H$ affect the number of blocks that can reside on each SM (Table~\ref{tab:arch}).

\begin{figure*}[t]
\centering
\includegraphics[width=\textwidth]{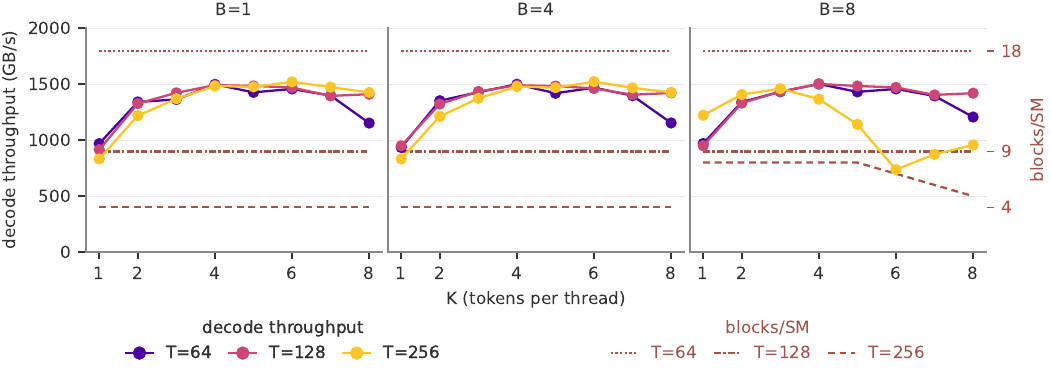}
\caption{Decode throughput and blocks/SM when varying $K$, across multiple combinations of launch bounds $(T=\{64, 128, 256\}, B=\{1,4,8\})$. B300, Loghub \texttt{Windows}, OnPair-12 ($W=8$, $H=1$).}
\label{fig:grid}
\end{figure*}

First, each SM has a fixed number of warp slots, $N_w$. A block of $T$ threads occupies $T/32$ of them, one per warp, so at most $\lfloor N_w/(T/32)\rfloor$ blocks are ever resident. The three HBM-based devices in our evaluation set $N_w=64$, while the GDDR ones set $N_w=48$ (Table~\ref{tab:arch}).

Next, each SM has a fixed-size register file: $R = 65\,536$, with 32-bit registers, for every GPU we evaluate. Each SM-resident thread draws from this shared pool. The launch-bound parameter $B$ asks the compiler to budget registers for at least $B$ blocks of $T$ threads per SM~\cite{cudaguide}. This gives a per-thread register limit of approximately $R / (T \cdot B)$; increasing $B$ leaves fewer registers for each thread. Note that this compiler target does not guarantee that $B$ blocks fit under the other resource constraints.

Further, coarsening and hoisting both require register space. Each of a thread's $K$ codes requires registers for the code itself, the token's length, the output offset, and its low-plane bytes. Each hoisted round retains one high-plane value per lane. Using $L$ for the maximum token length supported by the kernel and $c$ for the per-thread overhead state, an approximate register-limit condition for a pair $(K,H)$ is:
\[
  K\!\left(3 + \frac{W}{4}\right) + H\,\frac{L - W}{4} + c \;\le\; \frac{R}{T \cdot B} 
\]

This construction allocates one register each for a code, its token length, and its offset. An additional $W/4$ registers store low-plane dictionary bytes, and $(L-W)/4$ store hoisted high-plane bytes. Note that at $W = L$, every token fits in the low plane, so there are no high-plane bytes or a request queue, eliminating the impact of $H$.

Finally, each block reserves shared memory for its warps' staging buffers and request queues. Increasing $K$ requires more output space and queue slots, increasing the per-block allocation $M$ (including overheads). If $A$ bytes of shared memory are available per SM, at most $\lfloor A/M\rfloor$ such blocks fit.

Each of the established constraints may limit how many blocks an SM holds at once (given $r$ registers per thread and $N_w$ warp slots per SM):

\[
  \min\!\left(
  \left\lfloor \frac{A}{M} \right\rfloor,\;
  \left\lfloor \frac{R}{T \cdot r} \right\rfloor,\;
  \left\lfloor \frac{N_w}{T/32} \right\rfloor \right)
\]

For example, at $T=256$, $B=4$, $K=6$, and $W=8$, the per-thread register limit is \claimRegsBfour{} registers. A block occupies \claimBlockKiB{}\,KiB, so four resident blocks require \claimFourBlocksKiB{}\,KiB. The three HBM GPUs make between 164 and 228\,KiB available, which is enough for all four blocks, while the two GDDR devices make only 100\,KiB available, which fits only three blocks.

Increasing the tokens decoded per thread ($K$) increases the required shared memory, which in turn can reduce the L1 cache available for the dictionary. At OnPair-12, the low plane is 32\,KiB for split reads ($W=8$); at OnPair-16, it is \claimLowPlaneSixteen{}\,KiB at the same width, larger than any evaluated GPU's total per-SM SRAM. The compiler can also add to this cache pressure, as registers it cannot keep spill to \emph{local memory}, per-thread device-memory storage cached like any other access; these spilled-register reads then compete with dictionary reads~\cite{cudaguide}.

In the next section, we explore the impact of these parameters through multiple sensitivity studies.

\section{Microbenchmarks}
\label{sec:mb}

This section addresses the following questions through a set of targeted experiments:

\begin{figure*}[t]
\centering
\includegraphics[width=\textwidth]{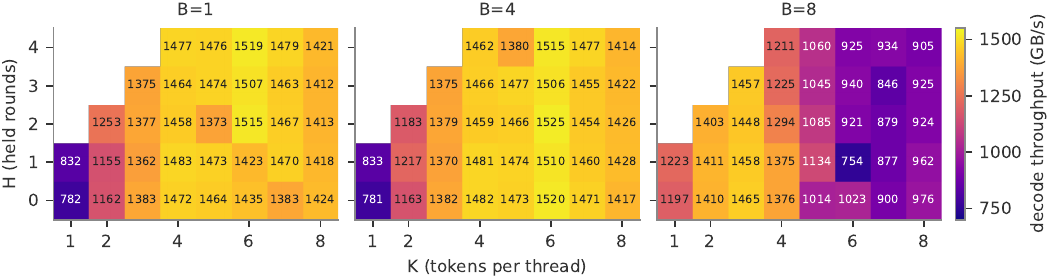}
\caption{Decode throughput when varying $(K,H)$. ($T=256, B=\{1,4,8\}$). B300, Loghub \texttt{Windows}, 
OnPair-12 ($W=8$).
}
\label{fig:hoist}
\end{figure*}

\begin{enumerate}
\renewcommand{\labelenumi}{\textbf{Q\arabic{enumi}}}
  \item How do the $K$ (tokens per thread), $T$ (threads per block), and $B$ (resident blocks requested) parameters impact FastPair's decode rate? (Section~\ref{sec:mb:launch})
  \item Does issuing the long-token reads early (the hoist stage) make decoding faster? (Section~\ref{sec:mb:hoist})
  \item Does unconditionally fetching only part of each dictionary entry result in a faster decoder? Is this benefit data-dependent? (Section~\ref{sec:mb:gatherwidth})
  \item Would staging the dictionary in shared memory result in faster decoding? (Section~\ref{sec:mb:shdict})
  \item How does mean token length affect decode rate? Does this depend on whether the dictionary fits in the L1 cache? (Section~\ref{sec:mb:width})
\end{enumerate}

Unless stated otherwise, rates are reported in GB/s using the minimum time from at least 100 decodes on a B300, at its boost clock.

\subsection{Kernel Launch Parameters}
\label{sec:mb:launch}

To better understand decoder parallelism limits, we examine the effects of $K$ (tokens per thread) and the launch bounds $T$ (threads per block) and $B$ (resident blocks requested). We sweep these parameters by decoding the Loghub \texttt{Windows} column on a B300 using OnPair-12 ($W=8$, $H=1$). The results of this study are shown in Figure~\ref{fig:grid}.

As $K$ increases, each thread performs more work and requires more registers. We observe similar behavior at $B=1$ and $B=4$, whereas $B=8$ behaves differently. At $T=256$, increasing $B$ from 4 to 8 halves the register limit to \claimRegsBeightLaunch{} per thread. For these parameters, after $K=5$, the kernel spills registers to local memory and reduces residency from eight blocks to five. These spills add L1 accesses that compete with dictionary reads; thus, even though more blocks can reside than at $B=4$, this kernel decodes more slowly. 

Based on these experiments, we recommend $T=256$, $B=4$, and $K=6$ as the default launch configuration on the B300, as it achieves the highest decode rate in the sweep and provides capacity for four blocks per SM without spilling.

\subsection{Hoisted Reads}
\label{sec:mb:hoist}

To evaluate the impact of hoisting reads, we perform a sensitivity study by decoding the Loghub \texttt{Windows} column using OnPair-12 ($W=8$) on a B300, varying $K$ (tokens per thread) and $H$ (held rounds of hoisted reads) across launch configurations ($T=256, B=\{1,4,8\}$). The results of this experiment are shown in Figure~\ref{fig:hoist}.

The hoist does not improve the highest-performing configuration ($T=256, B=4, K=6$): at $K=6$ the coarsened loop already has enough independent loads in flight to hide their latency (Section~\ref{sec:mb:launch}). The hoist matters at $(B=\{1,4\}, K=1)$, where too few loads are outstanding to hide the latency of any one of them. Issuing the high-plane reads before the emit stage leaves fewer warps stalled on an outstanding load.\footnote{At $(T=256, B=1, K=1)$, where $H=1$ improves on $H=0$ by 6.4\%, the share of warp cycles stalled on an outstanding L1 load falls from 39.4\% to 33.2\%, while the shared-memory and memory-queue stalls rise.} We use $H=1$ by default, which holds one round, and drop to $H=0$ only when the compiler reports a register spill.

Register utilization also explains why hoisting degrades performance under the tighter register limit at $B=8$, where, at $K=4$, increasing $H$ to two triggers spilling and reduces performance. Here, the additional register pressure causes registers to spill to local memory. Because local memory traffic is cached in L1, spilled registers compete with the dictionary. %

\begin{figure}[b]
\centering
\includegraphics[width=\columnwidth]{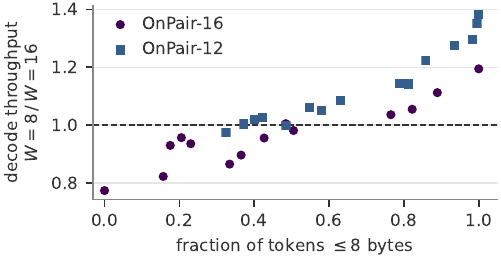}
\caption{Ratio of the best decode throughputs found for $W=8$ and $W=16$, per column. Columns are organized by their fraction of tokens $\leq 8$ bytes. B300, OnPair-16 and -12. 
}
\label{fig:gatherwidth}
\end{figure}

\subsection{Gather Width}
\label{sec:mb:gatherwidth}

Splitting decoding dictionaries into a low plane and high plane lets a single indexed load serve every token of at most $W$ bytes, while longer tokens require a second (high-plane) lookup. This optimization enables narrower reads, which may require fewer L1 wavefronts even when the same number of sectors are accessed (Section~\ref{sec:bg:gpu}). The benefit of this split therefore depends on the data: specifically, the portion of a column's tokens the low plane can serve on its own. We show the results of exploring $W$ in Figure~\ref{fig:gatherwidth}.

The narrow read ($W=8$) is slower than the full read ($W=16$) when long tokens are common. An extreme example is the \texttt{o\_clerk} column decoded by OnPair-16, where no token is eight bytes or fewer, and the narrow read is 23\% slower. The narrow read begins to outperform the full read once the low plane serves more than about 40\% of a column's tokens.
The most beneficial columns for the narrow read, using OnPair-12, are \texttt{c\_address} and the FineWeb2 Mandarin columns, which decode 1.38$\times$ and 1.35$\times$ faster, respectively. This behavior supports automatic kernel selection, since the encoder already holds the token-length histogram when it writes the dictionary. We recommend defaulting to split reads ($W=8$) and disabling them ($W=16$) when over 60\% of a column's tokens exceed eight bytes. 

\subsection{Shared-Memory-Resident Dictionaries}
\label{sec:mb:shdict}

Prior GPU-based FSST-decoding work stages the dictionary in shared memory, eliminating the global-memory gather~\cite{gsst}. While this approach works well for FSST's small dictionary, larger dictionaries pose two challenges: First, they may not fit in shared memory. Second, staging a dictionary in each thread block consumes shared memory, potentially limiting the number of resident blocks. We therefore measure the impact of staging the decoding dictionaries in shared memory when decoding the Loghub \texttt{Windows} column using a B300.

OnPair-16's low plane alone is 512\,KiB, so it cannot fit in the B300's shared-memory carveout (228\,KiB; Table~\ref{tab:arch}). Where staging does fit, it is slower: by 26\% at OnPair-12 (1131\,GB/s against 1527\,GB/s) and by 54\% at FSST-12 (643\,GB/s against 1409\,GB/s).

We investigate performance counters to better understand the impact of staging the dictionary in shared memory, profiling the best staging kernel variant against the non-staging kernel for this column. At OnPair-12, staging cuts global load sectors from 209 million to 7.6 million per launch, and the sectors each request touches from 17.4 to 1.9, indicating that the global-memory gather is gone. However, shared-memory bank conflicts rise from 61 million to 91 million, occupancy falls from 52\% to 24\%, and the L1 pipe falls from 99\% to 76\% of its access-rate limit rather than being freed. Staging replaces global-memory dictionary reads with shared-memory reads, which still compete with output writes for the L1 pipe's access rate (Section~\ref{sec:eval:bound}). We therefore leave larger dictionaries to the GPU's existing cache structures.

\subsection{Token Length and Decode Rate}
\label{sec:mb:width}

\begin{figure}[t]
\centering
\includegraphics[width=\columnwidth]{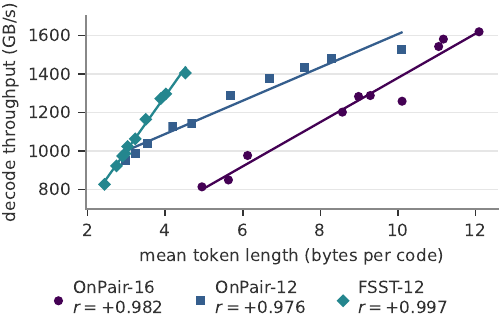}
\caption{Decode throughput against mean token length on a B300, with least-squares fits and correlation coefficients.
}
\label{fig:lenpredict}
\end{figure}

Across the real-world data columns, the decode rate rises nearly linearly with mean token length (Figure~\ref{fig:lenpredict}).

The number of bytes decoded per second equals the mean token length multiplied by the number of codes decoded per second. The compression ratio equals the mean token length divided by the mean bytes per code (ignoring the dictionary, sidecar, and further cascaded compression). At a fixed code decoding rate and code width, increasing the mean token length therefore improves both decode throughput and the compression ratio.
FSST-family tokens are decoded independently, regardless of length. A higher mean token length means fewer codes for the same text, improving both compression ratio and decoding throughput~\cite{onpair}.

Further, this decoding relationship does not require the dictionary to be L1-resident. OnPair-16's low-plane dictionary at $W=8$ is \claimLowPlaneSixteen{}\,KiB, at least twice as large as any explored GPU's maximum per-SM SRAM capacity, so it cannot be L1-resident on any of those GPUs. OnPair-16 remains the fastest decoding technique on five of the ten real columns.

\section{Evaluation}
\label{sec:evaluation}

\begin{figure*}[t]
\centering
\includegraphics[width=\textwidth]{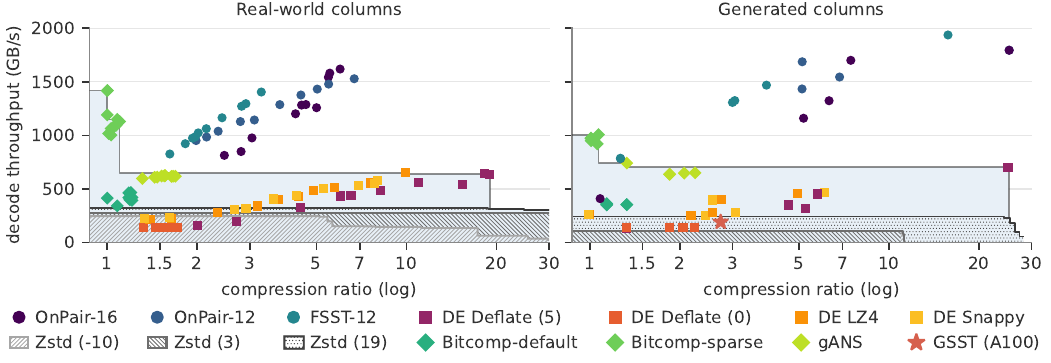}
\caption{Decode throughput against compression ratio on a B300, for both real-world and generated data columns. 
We show a baseline envelope: at each compression ratio, the fastest rate any baseline technique reaches at that ratio or better. We also include GSST's A100-based result for reference. %
Zstd, at each compression level, is drawn as a best-found envelope.
}
\label{fig:perf_real}
\end{figure*}

We evaluate FastPair on the five GPUs listed in Table~\ref{tab:arch}, over the first 1\,GB of each column in Table~\ref{tab:datasets}~\cite{fineweb2,wikidump,codeparrot,loghub,clickbench,tpch}. We compute decode rates from the minimum decoding time across at least 100 runs, with clock boosting enabled, unless specified otherwise.

We prepare all relevant data and move it to device memory before decoding begins. In particular, we widen the stored, bit-packed codes and repack the dictionary into fixed-stride entries. We also decompress the offset sidecar and de-cascade the FSST-family encodings to their base form.

We emphasize that the compression ratios reported throughout this work refer to data organized as a variable-length string array, where the string data is stored byte-packed and is accompanied by an array of fixed-width offsets. For each technique, we compress the string data using the technique under evaluation. Because decompressing the row offsets array is not necessary for bulk decompression, we do not include it in the compression ratio. Thus, for general-purpose compression techniques, the compression ratio is the ratio of uncompressed to compressed bytes, while for FSST-family codecs we report uncompressed bytes against the sum of cascade-compressed bytes, batch offsets (sidecar), and dictionary data.

\begin{figure}[b]
\centering
\includegraphics[width=\columnwidth]{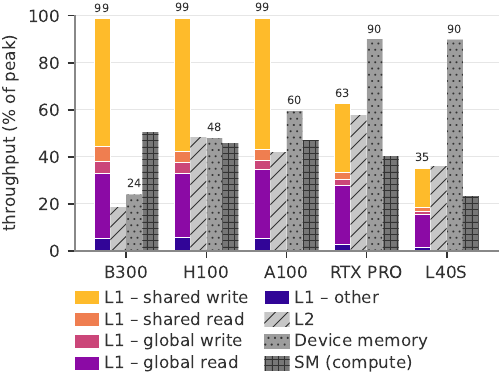}
\caption{
Unit throughput as a fraction of its peak sustained rate (boost enabled)~\cite{ncu}. Loghub \texttt{Windows}. OnPair-12. %
}
\label{fig:pipes}
\end{figure}

Broadly, we recommend configuring FastPair with $K=6$ (tokens per thread), $H=1$ (held rounds of hoisted reads), and split reads enabled based on the column's token-length distribution. For the B300, we also recommend launch bounds $T=256$ (threads per block) and $B=4$ (resident blocks requested). However, unless stated otherwise, the results in this section report each technique, both ours and each baseline, at the best configuration we found for each column. Where configurations trade compression ratio for decoding speed, such as Zstd frame-size tuning, we show an envelope of the best configurations found.

\subsection{Performance Evaluation}
\label{sec:eval:perf}

\begin{figure*}[t]
\centering
\includegraphics[width=\textwidth]{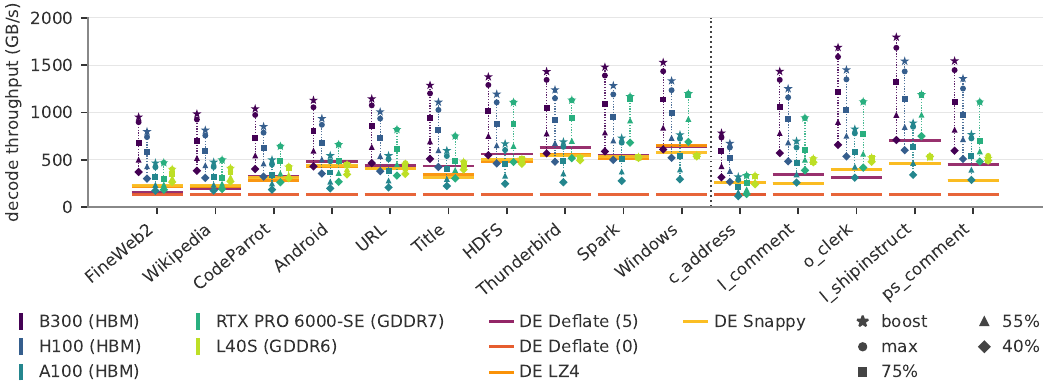}
\caption{Decode throughput of OnPair-12 on five GPUs across five SM clock frequencies: unlocked (boost), locked at max, and locked at 75, 55, and 40\% of its maximum. 
The horizontal lines are the fastest configurations of the four B300 DE settings.
}
\label{fig:perf_gen}
\end{figure*}

On the B300, FastPair decodes the \claimRealColumns{} real-data columns at \claimBthreeRealMinTwelve{}--\claimBthreeRealMaxTwelve{}\,GB/s using OnPair-12 and \claimBthreeRealMinSixteen{}--\claimBthreeRealMaxSixteen{}\,GB/s using OnPair-16. The fastest column is \claimFastestColumn{} at \claimFastestColumnRate{}\,GB/s using \claimFastestColumnPreset{}, against \claimFastestColumnDeRate{}\,GB/s for the DE's best codec and chunk size on that same column; an H100 decodes it at \claimFastestColumnHopperRate{}\,GB/s, above the B300's Decompression Engine. 
Among the real-world columns, no baseline configuration we measured reaches those rates at an equal or better compression ratio. These results are shown in Figure~\ref{fig:perf_real}.

The only technique that shows comparable decoding performance on real-world data is the nvCOMP Bitcomp kernel using the Sparse compression mode; however, this kernel targets scientific data (floating-point) and barely compresses the evaluated string-data columns. Among the baselines that compress these columns by at least 2$\times$, the DE decodes fastest on every real-world column; every baseline that reaches this minimum compression ratio uses a back-reference-based codec, the kind the DE accelerates.

Further, neither OnPair-12 nor OnPair-16 is strictly better than the other. OnPair-16 is faster on four of the five log columns and slower on the
text corpora, by as much as \claimPresetFlipWorst{} across the devices we measure. Thus, the choice between OnPair-12 and -16 is largely data-dependent; if a given column can use the larger dictionary and regularly forms tokens longer than eight bytes, such as the aforementioned log columns with repeated structural components, then OnPair-16 may outperform OnPair-12.

\subsection{Performance Analysis}
\label{sec:eval:bound}

To identify the factors that limit FastPair's decoding throughput, we use device performance counters to measure unit utilization. We express each unit's throughput relative to its peak sustained rate and derive the breakdown of the L1 pipe's wavefront activity, which we show in Figure~\ref{fig:pipes}.

On the B300, H100, and A100, the per-SM L1 pipe reaches about 99\% of its access-rate limit when decoding the Loghub \texttt{Windows} column using OnPair-12, while device-memory bandwidth utilization ranges from 24\% to 60\% of peak. This contrasts with the RTX PRO 6000-SE and L40S, which reach 63\% and 35\% of that limit, respectively, while using about 90\% of device-memory bandwidth. On the three HBM devices, the L1 pipe is nearly saturated while device-memory bandwidth remains available.

Shared writes account for roughly 56\% of the L1 pipe's wavefronts on the HBM devices and 47\% on the GDDR devices. These writes include output assembly in the emit stage and request-queue writes in split-read kernels. Global reads, including dictionary lookups in the gather stage, form the second-largest component. The remainder is other shared-memory traffic, including the warp shuffles used by the scan. The two largest consumers of the L1 pipe are bandwidth-inefficient: scattered dictionary reads and short shared-memory writes consume access capacity while transferring little data. In contrast, the drain stage combines neighboring threads' accesses into more efficient coalesced transfers, reading from shared memory to write to global memory.

\subsection{Cross-Architecture Evaluation}
\label{sec:eval:arch}

Using the B300's Decompression Engine as the baseline, we compare FastPair's OnPair-12 decoding performance across the five GPUs. We also vary each GPU's SM clock, while keeping the memory clock fixed, across five states: unlocked (boost), locked at max, and locked at 75, 55, and 40\% of its maximum. These results are shown in Figure~\ref{fig:perf_gen}.

On the B300, FastPair decodes every column we explore faster than the DE. Older or less provisioned GPUs also exceed it on most columns: the H100, A100, and RTX PRO 6000-SE decode faster than the DE on \claimBeatsDeTrioPhrase{} real columns, while the GDDR6 L40S does so on only \claimBeatsDeLfortySPhrase{}, falling short on \claimShortLoghubLfortySPhrase{} Loghub columns.

FastPair's decoding performance on the B300, H100, and A100 scales nearly linearly with the SM clock. On these devices, columns with longer tokens decode faster and show larger absolute throughput gains as the clock increases. The RTX PRO 6000-SE levels off near the top of its clock range, particularly on long-token log columns.
On the L40S, while longer tokens also improve decoding throughput, they reduce the impact of higher SM clocks.

\section{Related Work}
\label{sec:related}

GSST~\cite{gsst} is the closest prior GPU decoder for FSST-family strings. At compression time, GSST divides input into blocks with separate FSST-8 dictionaries, then subdivides each block into independently decodable splits. Splits are constructed to share either a compressed or uncompressed size; GSST stores the other size and recovers starting offsets with a prefix sum. GSST decodes the data by assigning each split to a thread. Both GSST and FastPair use a staged-then-aligned drain~\cite{gsst} and both give each thread many tokens rather than one~\cite{volkov}.
In contrast to GSST, decoding escape-free codes enables FastPair to construct batches without changing the code sequence. Further, its larger dictionaries preclude GSST's approach of always storing the entire dictionary in shared memory. Finally, FastPair distributes a batch's codes across a warp for decoding.

DPF~\cite{dpf} decodes FSST records in parallel, using a prefix sum for output placement and shared-memory staging for coalesced writes. %
Fang et al.~\cite{fang2010} position variable-length decoded output with a parallel prefix sum and stage small dictionaries in shared memory.
CODAG~\cite{codag} uses warp-level decoding and coalesced I/O for RLE and Deflate. Its lanes repeat the same serial decoding work to avoid broadcasting decoder state, whereas FastPair assigns independent dictionary lookups to lanes.  ZipFlow~\cite{zipflow} generates GPU kernels for string-dictionary expansion, including a word dictionary cascaded with bit-packing and ANS.

\section{Conclusion and Future Work}
\label{sec:conclusion}

FastPair decodes FSST-family compressed strings efficiently on GPUs by reorganizing dictionary lookups and output writes. Across ten real-world columns on a B300, its best per-column configuration achieves 2.4 to 4.2$\times$ the decoding rate of the fixed-function Decompression Engine. Our architectural study shows that dictionary reads and the short shared-memory writes used to assemble output can reach the L1 pipe's access-rate limit. %
Decoding throughput therefore depends on the memory accesses required per decoded byte, as well as on the parallelism exposed by the codec. 

FastPair contributes to a broader body of work built around cascading lightweight encodings to compress data. In future work, we aim to explore other lightweight encodings and extend our techniques to other compute platforms.

\section*{Acknowledgments}

Generative AI assisted with the development and evaluation of FastPair and the preparation of this manuscript. The authors directed all use of AI and stand by this work.

\bibliographystyle{ACM-Reference-Format}
\bibliography{fastpair}

\end{document}